\documentclass[prd,superscriptaddress,nofootinbib,showkeys, preprint]{revtex4}
\usepackage{graphicx}
\usepackage[usenames,dvipsnames]{color}
\usepackage{amsmath,amssymb}
\usepackage{mathtools}
\usepackage[colorlinks]{hyperref}
\usepackage{float}

\newcommand{\be}{\begin{eqnarray}}                                             
\newcommand{\ee}{\end{eqnarray}}
\newcommand{\nn}{\nonumber}

\newcommand{\noplus}{}

\newcommand{\tmop}[1]{\ensuremath{\operatorname{#1}}}
\begin{document}
	\title{ Deconfinement to Confinement as PT phase transition}
		
	\author{Haresh Raval}
	\email{haresh@phy.iitb.ac.in} 
	\affiliation{Department of Physics, Institute of Science, Banaras 
		Hindu University, 
		Varanasi - 221005, India}
	\affiliation{ Department of Physics, Indian Institute of Technology Delhi, New Delhi-11016, India}
	\author{Bhabani Prasad Mandal}
	\email{bhabani.mandal@gmail.com}
	\affiliation{Department of Physics, Institute of Science, Banaras 
		Hindu University, 
		Varanasi - 221005, India}
		\begin{abstract}
		
		We  consider $SU(N)$ QCD in a new quadratic gauge which highlights certain characteristic of the theory in the non-perturbative sector. By considering natural hermiticity property of the ghost fields we cast this model as non-Hermtian but symmetric under  combined Parity (P) and Time reversal (T)  transformations. We explicitly study the PT phase transition in this model. This is very first such study in the non-Hermitian gauge theory.  The ghost fields condensate  which give rise to spontaneous breaking of  PT symmetry. This leads to realize the transition from deconfined  phase to confined phase as a PT phase transition in this system. The hidden C-symmetry in this system is identified as inner automorphism in this theory. Explicit representation is constructed for the C-symmetry.
		
		\end{abstract}
	
	\maketitle
	
	\section{Introduction}
	Symmetries and their spontaneous break down played crucial role in the understanding  of physics from time to time. About two decades ago the  formulation of usual quantum mechanics where all physical observables are represented by self adjoint operators, has been extended to include non-self adjoint operators for their observable. Two important discrete symmetries,
namely Parity (P) and Time Reversal (T) are instrumental in such formulation as first shown in Ref.~\cite{11,12}.  Consistent  formulation with real energy eigenvalues, unitary time evolution and probabilistic interpretation for unbroken PT symmetric non-Hermitian quantum systems  have been formulated in a different Hilbert space equipped with positive definite CPT inner product. Such non-Hermitian PT symmetric systems generally exhibit a phase transition  ( or more specifically a PT breaking transition ) that separates two parametric regions: (i) region of the unbroken PT symmetry in which the entire spectrum is real and eigenstates of the systems respect PT symmetry and (ii) a region of the broken PT symmetry in which the whole spectrum (or a part of it) appears as complex conjugate pairs and the eigenstates of the systems do not respect PT symmetry.
The physics at this transition point is extremely rich in nature and the typical characteristics  of the non-Hermitian system are reflected by the behavior of the system at the transition point.
Thus, the  PT phase transition  and its realization  being extremely important in theories with non-hermitian systems have been studied frequently~\cite{nh1,nh2,nh3,nh34,nh35,nh36,nh4,nh5,nh6,nh8,nh9,nh10}. Even though self-adjointness  of quantum observables have never been challenged, this formulation of complex quantum mechanics created remarkable interest  in several fronts of physics including open quantum systems~\cite{oq1}, scattering theory~\cite{cs1,cs2,cs3,cs4,cs5,cs6,cs7,cs8,cs9}, optics etc. Particularly the theory of optics where several physical processes are known to follow Schrödinger like equations, provides a fertile ground  to verify the implications of such formulations experimentally. PT symmetric complex potentials are realized through complex refractive index in the optical media and the consequence of PT phase transition has been experimentally realized in optics~\cite{nh10,cs1,o1,o3,o4}. Therefore,
the applicability of this path breaking formulation  of complex quantum theories  relies on  observing PT phase transition in various physical systems.

In the present work, we for the first time demonstrate the PT phase transition in a  gauge theory.  We consider $SU(N)$ QCD in the newly found quadratic gauge \cite{3}, which is shown recently to have substantial  implications in the non-perturbative sector of the theory~\cite{3,4},  to study the PT phase transition. This is a novel study of PT phase transition in a gauge theory. Although the non-Hermitian extension of a gauge theory has been explored~\cite{ng1,1}, this particular subject has never been touched upon. The gauge has a few following unusual features. (1) The gauge is not one of Abelian projection gauges~\cite{gt} and has quark confinement signatures contrary to common studies of the confinement which have been done in Abelian projection. (2) It is the covariant algebraic gauge. In general, algebraic gauges are not covariant. (3) It removes the Gribov ambiguity on the compact manifold contrary to the case of usual gauges.~\cite{4}. This theory has two distinct phases, one is normal phase or deconfined  phase and  in the other ghost fields condensate leading to the confinement phase~\cite{3}.  
The Lagrangian density which represents the deconfinement phase of the theory is shown to be non-Hermitian  by adopting the natural but unconventional property of  hermiticity for ghosts~\cite{kugo}. However, the theory is invariant under PT transformations of the gluon and ghost fields. We explicitly show that the appearance of the ghost condensed state is the cause of  spontaneous break down of the PT symmetry. At this transition point the theory passes from deconfined phase to confined phase. We further identify the inner automorphism in this system as  the  C-symmetry which is inherent in all PT symmetric systems and connects the negative PT norm states to positive PT norm states and vice-versa. This C-symmetry is useful to define the non-Hermitian theory in a fully consistent manner in the modified Hilbert space endowed with CPT inner product. We explicitly construct the representation of the C-symmetry for the present non Abelian theory. This provides us the first example of the explicit representation of the C-symmetry in any gauge theory. Hence, the present theory can be viewed as consistent non-Hermitian gauge theory.

Now we present the plan of the paper.
In the next section, we consider a charged scalar theory with non-Hermitian mass matrix to set the mathematical preliminaries for the later non Abelian non-Hermitian model. Two phases of the QCD  in the newly found quadratic gauge have been elaborated in Sec. III. In Sec. IV, non hermiticity  and PT symmetry properties of two phases  have been discussed. The transition from deconfined phase to confined phase has also been identified as  PT phase transition in this section. Explicit representation of the C-symmetry is constructed. Last Sec. is kept for results and discussions.

\section{The toy model of non Hermitian complex scalars}	
	In what follows from the next section has a close analogy with simple complex scalar non-Hermitian model discussed in Refs.~\cite{1,2}. Hence, we first study the  non-Hermitian theory of charged scalars.
	This theory is described by the following Lagrangian
	\be
	L = \partial_\mu \phi_1^*\partial_\mu \phi_1 +\partial_\mu \phi_2^*\partial_\mu \phi_2 + [\phi_1^* \ \  \phi_2^*]M^2
	\begin{bmatrix*}
		\phi_1  \\ \phi_2
	\end{bmatrix*}
\ee
where \be M^2 = \begin{bmatrix*}
	m_1^2 & \mu^2\\
	-\mu^2 & m_2^2. 
\end{bmatrix*}
\ee
We will be interested only in cases for which 
$m_1^2, m_2^2, \mu^2 \geq 0$. We see that the mass matrix $M^2$ is not Hermitian. Discussion on discrete symmetries become easier when the doublet of two fields is defined as
\be
\Phi = \begin{bmatrix*}
	\phi_1  \\ \phi_2
\end{bmatrix*}.
\ee
Then, the parity  and time reversal respectively are defined on the doublet as follows
\be
\Phi  &\xrightarrow{\text{\ \ P\ \ }} &P\Phi\\
\Phi  &\xrightarrow{\text{\ \ T\ \ }}& T\Phi^*
\ee
where P and T now are $2\times2$ matrices and complex conjugation in time reversal is due to anti-linearity. We can make a clear guess for the choice of P by the analogy of the parity transformation in $\mathbb{R}^2$, where $ x \rightarrow x$ and $y \rightarrow -y$. The parity in $\mathbb{R}^2$ suggests  that   the field $\phi_1$  transforms as a scalar  and the other, $\phi_2$ transforms as a pseudo scalar. Therefore, the P has the following matrix form
\be P=
\begin{bmatrix}
	1 & 0\\
	0 & -1.\\
\end{bmatrix}\ee
This leaves us with the only  choice for the time reversal T under which the Lagrangian is PT-invariant. We must choose T $= \mathbf{1}_2$~\cite{1,2}.
 One can in principle swap the roles of P and T however in order to interpret  
 this PT-symmetric theory in terms of a coupled system with gain and loss, one should
take T $= \mathbf{1}_2$~\cite{1,2}. The theory remains in the unbroken PT- symmetric state as long as the eigenvalues of the mass matrix given as below remain real,
\be\label{pm}
M_\pm^2 = \frac{1}{2}(m_1^2+ m^2_2)\pm \sqrt{(m_1^2- m^2_2)^2 - 4\mu^4}
\ee
So, for $|m_1^2-m_2^2| \geq2\mu^2$, we are in the phase of unbroken PT symmetry.
 When $|m_1^2-m_2^2| < 2\mu^2$ happens, we step into the region of broken PT-symmetry as eigenvalues turn complex and PT$ \psi_\pm = \pm \psi_\pm$ is no longer valid, where $\psi_\pm$ are eigenfunctions of the mass matrix $M^2$ corresponding to eigenvalues $M_\pm^2$. We shall  encounter similar non Hermitian mass matrix for gluons in our non Abelian model to be discussed.  

Since the eigenvalues in Eq.~\eqref{pm} do not change under $\mu^2 \rightarrow -\mu^2$, there still exist the charge conjugation symmetry under which the theory is CPT invariant in both PT broken and unbroken phases.  The  charge conjugation is defined as follows
\be
\Phi  &\xrightarrow{\text{\ \ C\ \ }} &C\Phi^*
\ee
with C=P~\cite{1,2}. The theory in the region $|m_1^2-m_2^2| < 2\mu^2$  violates CP also but preserves CT symmetry. Such charge conjugation symmetry  exist in the non Abelian model also as we will see later. 

	\section{SU(N) QCD in the quadratic gauge}
	Here we discuss a model in which we intend to establish a PT phase transition.  
 The model relies on the new type of quadratic gauge fixing of Yang-Mills action as follows~\cite{3},
	\begin{align} \label{eq:0}
	H^a [ A^{\mu} ( x) ] =
	A^a_{\mu} ( x) A^{\mu a} ( x) = f^a ( x) ; \  \text{  for each $a$ }
	\end{align}
	where $f^a(x)$ is an arbitrary function of $x$. 
	The	Faddeev-Popov determinant in this gauge is given by
	\be 
	\det\left(\frac{\delta (A^{a \epsilon}_\mu A^{\mu a \epsilon})}{\delta \epsilon^b}\right)= \det\left(2A^a_\mu( \partial^\mu \delta^{ab}-g f^{acb}A^{\mu c})\right),
	\ee
	Therefore,
	the resulting effective Lagrangian density is given as follows,
		\begin{equation} \label{LQ}
	\mathcal{L}_{Q}  =- \frac{1}{4} F^a_{\mu \nu} F^{\mu \nu a}
	\noplus - \frac{1}{2 \zeta}  ( A^a_{\mu} A^{\mu a})^2 - 2\overline{c^a}
	A^{\mu a} ( D_{\mu} c)^a  
	\end{equation}
	where $ \zeta$ is an arbitrary gauge fixing parameter, the field strength $F^a_{\mu \nu}= \partial_{\mu}A^a_{\nu}(x)- \partial_{\nu}A^a_{\mu}(x)-g 
	f^{abc} A^b_{\mu}(x)A^c_{\nu}(x)$ and $(D_{\mu} c)^a = 
	\partial_\mu c^a - g f^{a b c} A_\mu^b c^c$. 
The  summation over an index $a$ is 
	understood  when it  appears repeatedly, including when occurred thrice in the ghost term.
	In particular it is important for the present paper to understand a structure of the ghost Lagrangian,
	\begin{equation*}
	- \overline{c^a} A^{\mu a} ( D_{\mu} c)^a 
	= - \overline{c^a} A^{\mu a} \partial_{\mu} c^a + g f^{a b c} 
	\overline{c^a} c^c A^{\mu a} A^b_{\mu}
	\end{equation*}
	where the summation over indices $a$, $b$ and $c$ each runs independently over $1$ to $N^2-1$. 
	As shown in~\cite{4},
	The resulting Lagrangian is BRST  invariant\cite{Becchi:1974md, Becchi:1975nq}
	which is essential for the ghost independence of the green functions and unitarity of 
	the $S$-matrix. The substantial non-perturbative implications of this gauge have been studied in Refs.~\cite{3,4}.  In a recent in-
	teresting work, the FFBRST technique itself has been extended to connect non-perturbative
	sector implied by this gauge to perturbative sector signified by the Lorenz gauge~\cite{37}.
	
	\subsection{Phases of the theory in the quadratic gauge}
	This theory has two different phases~\cite{3}: the normal or deconfined phase and the ghost condensed phase showing the confinement. 
	 The Lagrangian in normal  phase is given by Eq.~\eqref{LQ} itself. We should note that the ghost Lagrangian does not have kinetic
	terms. They act like auxiliary fields in the normal phase, but play an important role in the IR regime as we discuss now.
	
	\subsection*{ Ghost condensation}
	To demonstrate the significance of ghosts in terms their condensates in the IR limit and citing its value for the present purpose also,  
	 we elaborate the ghost condensation and its implication thoroughly. The ghost condensation as a concept was  introduced independently in Refs.~\cite{new1,new2,new3}. 
	The  second term  in the ghost Lagrangian contains ghost bilinears multiplying terms quadratic in
	gauge fields. Hence if the ghosts freeze they amount to a non-zero mass matrix for the gluons as follows
	\begin{equation}
	(M^{ 2})^{a b}_{\tmop{dyn}} = 2 g \displaystyle\sum\limits_{c=1}^{N^2-1}f^{a b c} 
	\langle\overline{c^a} c^c\rangle.
	\end{equation}
	 	We would get masses of gluons  by  diagonalizing the matrix and finding its eigenvalues. In  
 an  $SU(N)$ symmetric state, where all ghost-anti{\small -}ghost 
	condensates are  identical as given in Eq.~\eqref{con1} below the mass matrix becomes peculiar, 
	\begin{equation}\label{con1}
	 \langle\overline{c^1} c^1\rangle = ... =  \langle\overline{c^1} c^{N^2-1}\rangle = ... =  \langle\overline{c^{N^2-1}} c^1\rangle = ... =     
	\langle\overline{c^{N^2-1}} c^{N^2-1}\rangle \equiv K.
	\end{equation}
	The physical relevance of the theory is lent strength once a physical mechanism is laid out within which the state in above equation can occur consistently.
	 This objective was achieved  by 
	introducing a Lorenz gauge fixing term for one of the diagonal gluons, in addition to the purely quadratic terms of Eq. \eqref{eq:0}.  
	This gauge fixing gives the propagator to the corresponding ghost field. Using this ghost propagator, one can give nontrivial vacuum values 
	to bilinears $ \overline{c^a} c^c $ within the framework Coleman-Weinberg mechanism as described 
	in \cite{3}. This mechanism naturally gives the K to be real.
	Thus, in the state of condensates given by Eq.~\eqref{con1}, the mass matrix becomes
	\begin{equation}\label{mm}
	(M^{ 2})^{a b}_{\tmop{dyn}} = 2 g  K\displaystyle\sum\limits_{c=1}^{N^2-1}f^{a b c} 
	\end{equation}
	which is an antisymmetric matrix i.e., non Hermitian, 
	\be
	(M^2)^\dagger  \neq M^2
	\ee
	due to the antisymmetry of the structure constants.
	 The matrix in Eq.~\eqref{mm} is unique, it has $N(N - 1)$ non-zero eigenvalues only and thus has 
	nullity $ N -1$ which implies that $N(N - 1)$ off-diagonal gluons obtain masses and the $N-1$ diagonal gluons remain massless. Because of the antisymmetry, eigenvalues occur are purely imaginary and  in conjugate pairs. 
	The massive off-diagonal gluons are inferred as evidence of Abelian dominance, which is one of  signatures of quark confinement. Further, mass squared of the off-diagonal gluon is purely imaginary, hence the pole of the off-diagonal gluon propagator is on imaginary $p^2$ axis which is another important signature of  color confinement~\cite{cd}. The mass for gluons generated through a given dynamical mechanism breaks the gauge symmetry as usual. We note that there exist other interesting mechanisms  where the mass can consistently be given to gluons in a gauge invariant manner, a thorough overview of such mechanisms is found in Refs.~\cite{new4, new5} and refs. therein.  Thus, we see that the ghost condensation acts as the QCD vacuum.  
	 	Therefore, in the ghost condensed phase the Lagrangian can effectively be given as follows
	 	\begin{eqnarray} \label{5}
	 	\mathcal{L}_{GC} =&-& \frac{1}{4} F^a_{\mu \nu} F^{\mu \nu a}
	 	\noplus - \frac{1}{2 \zeta}  ( A^a_{\mu} A^{\mu a})^2 + M^2_a A^a_{\mu} A^{\mu a}
	 	\end{eqnarray}
	 	Here for the diagonal gluons $M^2_a=0$, 
	 	e.g, for $SU(3)$, $M^2_3=M^2_8=0$. While for the off-diagonal gluons, $M^2_1=+im^2_1, M^2_2=-im^2_1, \  
	 	M^2_4=+im^2_2,  M^2_5=-im^2_2, \  M^2_6=+im^2_3, M^2_7=-im^2_3 \ (m_1^2, m_2^2, m_3^2 $ 
	 	are positive real$)$. So the  gluons 1 and 2 can be considered as conjugate of each other. The same is true for other pairs. Hence for $SU(3)$, the last term of the effective Lagrangian in Eq.~\eqref{5} would be
	 	\begin{eqnarray} \label{6}
	 	M^2_a A^a_{\mu} A^{\mu a}= & +&im^2_1 A^1_{\mu} A^{\mu 1}-im^2_1A^2_{\mu}A^{\mu 2} +im^2_2A^4_{\mu} A^{\mu 4}-im^2_2A^5_{\mu} A^{\mu 5} \nonumber \\&+&im^2_3A^6_{\mu} A^{\mu 6}-im^2_3A^7_{\mu} A^{\mu 7}
	 	\end{eqnarray}
	 	Thus we end our discussion on the quadratic gauge model.
	 	Having reviewed all the prerequisites, we are now in position to move on to the outlined objective of the work.
	 	
	 	\section{PT phase transition in the gauge theory}
	 	There have been studies on the non-Hermitian extension of a gauge theory. However, the subject of PT phase transition has not been explored in gauge theories.
	 	Here we show that the  non Abelian gauge theory  of interest exhibits the PT phase transition. Since discussion on PT symmetry becomes meaningful only in non-Hermitian systems we first discuss the hermiticity property of two mentioned phases and demonstrate that they both are non Hermitian.
	 	\subsection{Hermiticity of the theory}
	 	The effective Lagrangian in the normal phase is  given in Eq.~(\ref{LQ})
	 	\begin{eqnarray} 
	 	\mathcal{L}_{\tmop{eff}} &=&- \frac{1}{4} F^a_{\mu \nu} F^{\mu \nu a}
	 	\noplus - \frac{1}{2 \zeta}  ( A^a_{\mu} A^{\mu a})^2 - 2\overline{c^a}
	 	A^{\mu a} ( D_{\mu} c)^a
	 	\end{eqnarray}
	 	Now the hermiticity property of fields $A^a_\mu$ is well defined since they describe real degrees of freedom. Fields must be Hermitian in order to define the real degrees of freedom i.e., 
	 		\begin{eqnarray}\label{a}
	 	A_{\mu}^{a \dagger}&=& A_{\mu}^{a} 
	 	\end{eqnarray}
	 	However, such is not the case for ghosts. Their Hermiticity remains unclear. Based on the following heuristic argument, we shall define this property for ghosts under which the present theory is cast as non-Hermitian model. As the operation of conjugation in principle transforms particle to its anti particle, the following is the natural choice of Hermiticity property for ghosts\footnote[1]{In literature, at times un-conventional Hermiticity property   of ghost fields is invoked under which the  effective theory in a given usual gauge can be reinterpreted as Hermitian theory. However, this is not natural and methodological way of treatment.  (See Ref. [33]). The discussion on Hermiticity of Eq.~\eqref{LQ}  with natural Hermiticity Eq.~\eqref{c} is general to all usual gauges such as Lorenz gauge.}~\cite{kugo}
	 		\begin{eqnarray}\label{c}
	 c^{a \dagger}&=&\overline{c^{a}} \nonumber \\
	 	\overline{c^{a}}^ {\dagger}&=& c^{a}
	 	\end{eqnarray}
	 	Under Eqs.~\eqref{a},\eqref{c}, the Lagrangian in the normal phase in Eq.~(\ref{LQ}) is not Hermitian since the ghost Lagrangian is not Hermitian as shown below,
	 	\begin{eqnarray}
	 	&&(\overline{c^a} A^{\mu a} ( D_{\mu} c)^a)^\dagger 
	 	= (\overline{c^a} A^{\mu a} \partial_{\mu} c^a - g f^{a b c}
	 	\overline{c^a} c^c A^{\mu a} A^b_{\mu})^\dagger\nn \\ & =&- c^a A^{\mu a}   \partial_{\mu}\overline{c^a} + g f^{a b c} 
	 	c^a \overline{c^c} A^{\mu a} A^b_{\mu} \nn\\& =&  -c^a A^{\mu a} ( D_{\mu}  \overline{c^a})  \neq  \overline{c^a} A^{\mu a} ( D_{\mu} c)^a
	 	\end{eqnarray}
	 	Therefore,   the Lagrangian $(\mathcal{L}_Q)^\dagger$ can be written as 
	 		\be
	 	\mathcal{L}_Q^\dagger=	- \frac{1}{4} F^a_{\mu \nu} F^{\mu \nu a}
	 	- \frac{1}{2 \zeta}  ( A^a_{\mu} A^{\mu a})^2+ 2c^a A^{\mu a} ( D_{\mu}  \overline{c^a}).
	 	\ee
	 	Now, the Lagrangian $\mathcal{L}_Q$ is invariant under following BRST transformations
	 		\begin{eqnarray}\label{tra}
	 	\begin{split}
	 	\delta \overline{c^d }  =&  - \frac{\delta \omega}{g\zeta}A_\mu^d A^{\mu d} \\
	 	\delta c^d = &    \frac{\delta\omega}{2}f^{dbc} {c^b }  {c^c }  \\
	 	\delta A^d_\mu=&\ \frac{\delta \omega}{g}(D_\mu  {c^d }). \\
	 	\end{split}
	 	\end{eqnarray}  
	 		If the  Hermiticity properties of ghost fields are consistent with the BRST symmetry, then  $\mathcal{L}_Q^\dagger$ must be invariant under conjugate transformations we get when Hermitian conjugation (Eq.~\eqref{c}) is applied on both sides of transformations in Eq.~\eqref{tra} i.e.,  $( Eq.~\eqref{tra})^\dagger$ which are as follows
	 	\begin{eqnarray}\label{tra1}
	 	\begin{split}
	 	\delta c^{d \dagger}= &    \frac{\delta\omega}{2}(f^{dbc} {c^b }  {c^c })^\dagger  \Rightarrow	\delta \overline{c^d } =   - \frac{\delta\omega}{2}f^{dbc} \overline{c^b }  \overline{c^c }  \\
	 	\delta \overline{c^d }^\dagger  =&  - \frac{\delta \omega}{g\zeta}(A_\mu^d A^{\mu d} )^\dagger \Rightarrow	\delta c^d =-  \frac{\delta \omega}{g\zeta}A_\mu^d A^{\mu d} \\
	 	\delta A^{d \dagger}_\mu=&\ \frac{\delta \omega}{g}(D_\mu  {c^d })^\dagger \Rightarrow	\delta A^d_\mu=\ \frac{\delta \omega}{g}(D_\mu \overline{c^d }). \\
	 	\end{split}
	 	\end{eqnarray}
	 The Lagrangian $(\mathcal{L}_Q)^\dagger$ can be checked to be  invariant under  conjugate-BRST transformations of Eq.~\eqref{tra1} which supports firmly the physical relevance of the theory.
	 	   Therefore, the Hermiticity  properties of fields are consistent with the BRST symmetry of the theory. We further note that the Lagrangian  $\mathcal{L}_Q$ is very peculiar. It is only invariant under transformations in Eq.~\eqref{tra} and not under  Eq.~\eqref{tra1} and vice versa the Lagrangian   $\mathcal{L}_Q^\dagger$  is only invariant under  Eq.~\eqref{tra1}  and not under  Eq.~\eqref{tra}. 

		The effective Lagrangian in the ghost condensed (confinement) phase~\eqref{5} is also not Hermitian  as the  mass term for
 gluons is purely imaginary as explained. Important point here is to note that non hermiticity of the Lagrangian in this ghost condensed phase   is free of the hermiticity convention for ghosts as they do not appear in this phase and thus the non hermiticity of the ghost condensed phase is profound.
The Lagrangian \eqref{5} obeys the extended hermiticity~\cite{3} i.e., when the following inner automorphisms is applied hermiticity gets restored viz. $\mathfrak{T}\mathcal{L}_{GC}^{\dagger}\mathfrak{T}^{\dagger}=\mathcal{L}_{GC}$,
\begin{eqnarray}
\mathfrak{T}L_1\mathfrak{T}^{\dagger}=L_2 \hspace{1 cm} \mathfrak{T}L_4\mathfrak{T}^{\dagger}=L_5 \hspace{1 cm} \mathfrak{T}L_6\mathfrak{T}^{\dagger}=L_7 \hspace{1 cm} \mathfrak{T}L_3\mathfrak{T}^{\dagger}=L_8  \nonumber \\
\mathfrak{T}L_2\mathfrak{T}^{\dagger}=L_1\hspace{1 cm} \mathfrak{T}L_5\mathfrak{T}^{\dagger}=L_4\hspace{1 cm} \mathfrak{T}L_7\mathfrak{T}^{\dagger}=L_6 \hspace{1 cm} \mathfrak{T}L_8\mathfrak{T}^{\dagger}=L_3
\end{eqnarray}
with the property
\begin{align}
\mathfrak{T}^2= \mathfrak{T}^{\dagger 2}= 1 
\end{align}
where $L_i$ refers to any individual Lagrangian term such as $- \frac{1}{4} F^i_{\mu \nu} F^{\mu \nu i}$, $- \frac{1}{2 \zeta}  ( A^i_{\mu} A^{\mu i})^2$, $im^2 A^i_{\mu} A^{\mu i}$ appearing in Eq.~\eqref{5}.
The  inner automorphism amounts to exchanging group indices as $1\leftrightarrow 2$, $4\leftrightarrow 5$, $6\leftrightarrow 7$, $3\leftrightarrow 8$. 
In the adjoint representation it is given by

\begin{center}
\begin{equation}
	\begin{bmatrix}\label{mat}
	0 & 1 & 0 & 0 & 0 & 0 & 0 & 0\\
	1 & 0 & 0 & 0 & 0 & 0 & 0 & 0 \\
	0 & 0 & 0 & 0 & 0 & 0 & 0 & 1\\
	0 & 0 & 0 & 0 & 1 & 0 & 0 & 0 \\
	0 & 0 & 0 & 1 & 0 & 0 & 0 & 0\\
	0 & 0 & 0 & 0 & 0 & 0 & 1 & 0 \\
	0 & 0 & 0 & 0 & 0 & 1 & 0 & 0\\ 
	0 & 0 & 1 & 0 & 0 & 0 & 0 & 0 \\
	\end{bmatrix}
	\end{equation}
\end{center}
 We have thus shown that both deconfined and confined phases are non-Hermitian, later being  profoundly. There is no  spontaneous breaking of Hermiticity and the system is consistently non Hermitian. Hence, it becomes interesting to discuss state of PT symmetry in this theory which we shall commence now.
 \subsection{PT symmetry of the theory}
 As in the case of the hermiticity, parity and time reversal properties of the  gluons are well defined but not for ghosts. For gluons,
 parity is given as
 \be\label{par}
 A^a_i({\text{x},t}) \xrightarrow{\text{\ \ P\ \ }} - A^a_i(-{\text{x},t})\nn\\
 A^a_0({\text{x},t}) \xrightarrow{\text{\ \ P \ \ }}  A^{a }_0(-{\text{x},t}).
 \ee
 The rule for parity is same for all gluons as it is a linear operator. It is easy to see that Lagrangian in the normal phase~\eqref{LQ} is invariant  under parity if we choose  ghosts to be pseudo scalars or scalars. However, only  ghosts as scalars  are  consistent with the BRST transformations of fields under which the $\mathcal{L}_Q$ is invariant, 
  \be
 c^a({\text{x},t}) \xrightarrow{\text{\ \ P\ \ }} c^a(-{\text{x},t})\nn\\
 \overline{c^a}({\text{x},t}) \xrightarrow{\text{\ \ P \ \ }}  \overline{c^a}({-\text{x},t}).
 \ee
  Such convention is chosen in Ref.~\cite{qed}.  
 
The case of the time reversal is not straight forward unlike parity as the time reversal is an anti-linear operation. Since some of the generators of $SU(N)$ are purely imaginary, the time reversal property is not same for all gluons. We shall explain it using $SU(3)$ group, further generalization to $SU(N)$  is obvious. In $SU(3)$, three generators namely, 2nd,5th and 7th  are purely imaginary. Therefore, time reversal for gluons is given by
 \be
 A^p_i({\text{x},t}) &\xrightarrow{\text{\ \ T\ \ }}&   -A_i^p({\text{x},-t})\nn\\
 A^p_0({\text{x},t}) &\xrightarrow{\text{\ \ T \ \ }}&  A^{p }_0({\text{x},-t}),
 \ee
 where index $p$ is $1,3,4,6, 8$ and,
  \be
 A^q_i({\text{x},t}) &\xrightarrow{\text{\ \ T\ \ }}&   A^q_i({\text{x},-t})\nn\\
 A^q_0({\text{x},t}) &\xrightarrow{\text{\ \ T \ \ }}& - A^{q }_0({\text{x},-t}),
 \ee
 where index $q$ is $2,5,7$. Therefore, the field strength with any spacetime and group indices can utmost change up to overall negative sign i.e., 
 \be 
 F_{\mu \nu}^a  \xrightarrow{\text{\ \ T\ \ }} \pm F_{\mu \nu}^a.
 \ee
  Thus, the action in the normal phase \eqref{LQ} is invariant under time reversal  given that the time reversal property for ghosts is defined in the following manner, 
 \be
 c^p({\text{x},t})  \xrightarrow{\text{\ \ T\ \ }}  - c^p({\text{x},-t}) \nn\\
 \overline{c^p}({\text{x},t})  \xrightarrow{\text{\ \ T \ \ }}  \overline{c^p}({\text{x},-t}) 
 \ee and,
 \be
 c^q({\text{x},t})  \xrightarrow{\text{\ \ T\ \ }}   c^q({\text{x},-t}) \nn\\
 \overline{c^q}({\text{x},t})  \xrightarrow{\text{\ \ T \ \ }}  \overline{c^q}({\text{x},-t}) 
 \ee
 where the description of indices $p$ and $q$ are as above. Anti-linearity makes two sets of ghosts transform in a completely different manner.  It can be shown that these time reversal conventions are the only choices which are consistent with BRST transformations in Eq.~\eqref{tra}.  Thus,  the theory in normal phase is individually both parity and time reversal invariant.   This PT symmetry breaks down spontaneously in the confined phase as we explain now.
 
The theory in the confined phase is given by Eq.~\eqref{5},
\be
\mathcal{L}_{GC} =&-& \frac{1}{4} F^a_{\mu \nu} F^{\mu \nu a}
\noplus - \frac{1}{2 \zeta}  ( A^a_{\mu} A^{\mu a})^2 + M^2_a A^a_{\mu} A^{\mu a}\nn
\ee
 It is easy to check that parity~\eqref{par} is still a symmetry. However, the time reversal   is broken due to pure complex nature of the mass term,
\begin{eqnarray} \label{6}
M^2_a A^a_{\mu} A^{\mu a}= & +&im^2_1 A^1_{\mu} A^{\mu 1}-im^2_1A^2_{\mu}A^{\mu 2} +im^2_2A^4_{\mu} A^{\mu 4}-im^2_2A^5_{\mu} A^{\mu 5} \nonumber \\&+&im^2_3A^6_{\mu} A^{\mu 6}-im^2_3A^7_{\mu} A^{\mu 7}\ \ \xrightarrow{\text{\ \ T\ \ }}\ \ \\
& -&im^2_1 A^1_{\mu} A^{\mu 1}+im^2_1A^2_{\mu}A^{\mu 2} -im^2_2A^4_{\mu} A^{\mu 4}+im^2_2A^5_{\mu} A^{\mu 5} \nonumber \\&-&im^2_3A^6_{\mu} A^{\mu 6}+im^2_3A^7_{\mu} A^{\mu 7}\nn\\&=& -M^2_a A^a_{\mu} A^{\mu a}
\end{eqnarray}
 and also PT$\psi \neq \pm\psi$, where $\psi$s are eigenfunctions of the mass matrix~\eqref{mm}.  The first two terms of $\mathcal{L}_{GC}$ remain unaffected by the time-reversal. Thus, PT symmetry is violated in this phase. We can see that the anti symmetric nature of structure constant appearing in the mass matrix  has led to this breaking. Important point again here is to note that the PT symmetry violation in the confined phase is profound as it is independent of the convention for ghosts.  Therefore, the transition from the normal phase to the confinement phase with $SU(N)$ symmetric ghost condensates can be identified as PT phase transition from unbroken to broken PT phase. We  note here that interestingly association between color confinement and spontaneous PT breaking is model and mechanism independent even though in this model the link is through  ghost condensation since one prime signature of  quark confinement, the pole of the propagator on purely imaginary $p^2$ axis, inevitably breaks PT symmetry. Usefulness of a consistent model such as one in this paper lies in that it gives valuable insight into a process through which the link can take place.
  
   Conventionally the order parameter is the one whose value tuning separates two phases,  e.g.,  in the toy model of Sec. II,  tuning  of $\eta \equiv \frac{2\mu^2}{|m_1^2 -m_2^2|}$  from less than
  1 to greater than 1  separates PT unbroken and broken phase.   We note the following regarding an order parameter  in the present model. For the phase transition from deconfined phase which is PT unbroken to confined phase which PT broken, different ghost bilinears $\overline{c^a}c^c$ ($a$ and $c$ runs over $1$ to $N^2-1$ independently) need to condense first and that also to a stated $SU(N)$ symmetric vacuum given by Eq.~\eqref{con1}.  This mechanism is quite similar to the Higgs mechanism in the electroweak theory where ground state of theory parameterized by the expectation value of the Higgs field  spontaneously  breaks the  electroweak symmetry and thus in this sense, the  expectation value of the Higgs field acts as the order parameter. In the same way, the present theory has the ground state  parameterized by $K$ of Eq.~\eqref{con1} which   breaks the PT symmetry. Therefore, $K$ provides the order parameter of the PT transition in this non Abelian model. 
   Thus, we have provided a gauge theory in which PT phase transition is explicitly shown for the first time.
 
 \subsection{ C-symmetry}
 In the PT symmetric non Hermitian quantum mechanics, a C-symmetry (not the charge conjugation) is defined to improve the probabilistic interpretation of the PT-inner product and is inherent in all PT symmetric systems hence it becomes essential to find C-symmetry in the given model.  We show that in the setup of quantum field theory in which we are working the inner automorphism provides the representation of this C-symmetry. So far, no explicit representation of the C-symmetry is known in the framework of gauge theories. This symmetry in quantum mechanics must satisfy the following three conditions
 \be\label{con}
 [H, C]\psi=0, \ [PT, C]\psi=0, \ C^2=\bf{1}
 \ee
 The inner automorphism  satisfies QFT analogue of the conditions~\eqref{con} as we explain now.\\ (1) The inner automorphism exchanges 
 group indices i.e., $1\leftrightarrow 2$, $4\leftrightarrow 5$, $6\leftrightarrow 7$, $3\leftrightarrow 8$ and the Lagrangian of the initial unbroken PT theory in  the normal phase contains sum over group index $a$. Hence, the  Lagrangian  and therefore Hamiltonian in this  phase remain invariant under the inner automorphism. Thus,  QFT analogue of the first of conditions~\eqref{con} is obeyed.\\ (2) PT is a space-time symmetry and the inner automorphism is the operation in the group space. Therefore, it is easy to check that changing the order of inner automorphism and PT operations on Lagrangians of both the phases in Eqs.~\eqref{LQ} and  \eqref{5} does not alter the final result. In other words, they commute. This proves the QFT analogue of the second condition in Eq.~\eqref{con}.\\ (3) The third of Eq.~\eqref{con} has already been shown. 
 Therefore, we see that the inner automorphism forms an explicit representation of the  C-symmetry, which in adjoint representation is given by the matrix~\eqref{mat}.
 
 It is clear that the theory in both the phases is invariant under CPT. In the broken PT phase, the theory also violates CP symmetry but preserves the CT, in complete analogy with the scalar model described in sec. II. 
  
\section{ Conclusion}
	 	The main features of the non-self adjoint  theories are encoded in the rich characteristics of the PT phase transition, hence it is extremely important to study the PT phase in PT symmetric non self adjoint theories.  Even though  non-Hermitian extension of various  models in quantum field theory have been studied,  PT phase transition was not realized in the framework of a gauge theory. In the present work, we  have demonstrated the PT phase transition by constructing appropriate non-Hermitian but PT symmetric model of  QCD in a  recently introduced quadratic gauge which throws light  on 
		certain typical characteristics in non-perturbative sector. In this particular gauge, we have ghost fields condensation leading to confinement phase. We have shown the transition from deconfinement phase to confinement phase is a PT phase transition in this model of QCD. Ghost condensates  result in PT symmetry breakdown. To have a fully consistent non-Hermitian quantum theory,  it is important to explicitly find the C-symmetry which is inherent in all PT symmetric non-Hermitian systems. We have found the C-symmetry with its explicit representation in this model which is nothing but the inner automorphism. Thus, we give a new example where  representation of C-symmetry in a gauge theory is constructed. Importantly we  note that there is a direct association between quark confinement and PT breaking which we bring to light through this model for the first time. It would be interesting to further study the relevance that the implications of the PT-phase transition in this model may hold for the other areas of research.
		 
		\acknowledgments
		
		This work is partially supported by Department of Science and Technology, Govt. of India under National Postdoctoral Fellowship scheme
		with File no. `PDF/2017/000066'.


\begin{thebibliography}{39}
			\bibitem{11} C. M. Bender and Boettcher S, Phys. Rev. Lett. 80, 5243 (1998).
			\bibitem{12} C. M. Bender, Repts. Prog. Phys. 70, 947 (2007) and  Refs. therein.;A. Mostafazadeh, Int. J. Geom. Meth. Mod. Phys. 7, 1191(2010) and references therein.
			\bibitem{nh1} M. Znojil J. Phys. A 36 (2003) 7825.
\bibitem{nh2} B. P.Mandal, B. K Mourya, K.  Ali, A. Ghatak, Annals of Physics, Volume 363 (2015), 185-193.
\bibitem{nh3} C. M. Bender, S. Boettcher and P. N. Meisinger : J. Math. Phys. 40 2201 (1999).
\bibitem{nh34} A Khare, B P Mandal, Phys. Lett A, 272, 53 (2000).
\bibitem{nh35} B. P. Mandal, Mod. Phys. Lett. A 20, 655 (2005).
\bibitem{nh36} B. P. Mandal and A. Ghatak, J. Phys. A: Math. Theor. 45, 444022 (2012) .
\bibitem{nh4} C. T. West, T. Kottos, and T. Prosen: Phys. Rev. Lett. 104, 054102(2010).
\bibitem{nh5} A. Nanayakkara: Phys. Lett. A 304, 67 (2002).
\bibitem{nh6} C. M. Bender, G. V. Dunne, P. N. Meisinger, M. Simsek Phys. Lett. A 281 (2001)311-316.
\bibitem{nh8} B. P. Mandal, B.  K.  Mourya, R  K Yadav,  Physics Letters A 377, 1043 (2013). 
\bibitem{nh9} G. Levai, J. Phys. A 41 (2008) 244015. 
\bibitem{nh10}C. E. Ruter, K. G. Makris, R. El-Ganainy, D. N. Christodulides, M. Segev, and D. Kip: Nature (London) Phys. 6, 192 (2010).
       \bibitem{oq1} I. Rotter and J.P. Bird, Rep. Prog. Phys. 78, 114001 (2015) and Refs. therein.
  \bibitem{cs1} W. Wan, Y. Chong, L. Ge, H. Noh, A. D. Stone, H. Cao, Science 331, 889 (2011).
  \bibitem{cs2} N. Liu, M. Mesch, T. Weiss, M. Hentschel, and H. Giessen, Nano Lett. 10, 2342 (2010).
  \bibitem{cs3} H. Noh, Y. Chong, A. Douglas Stone, and Hui Cao, Phys. Rev. Lett. 108, 6805 (2011).
  \bibitem{cs4} C. F. Gmachl, Nature 467, 37 (2010).
  \bibitem{cs5} A. Ghatak, R. D. Ray Mandal, B. P. Mandal, Ann. of Phys. 336, 540 (2013).
  \bibitem{cs6} A Ghatak, Md.Hasan, B  P  Mandal, Phys. Lett. A 379, 1326 (2015).
  \bibitem{cs7} S. Longhi, J. Phys. A: Math. Theor. 44, 485302 (2011).
  \bibitem{cs8} RK Yadav, A Khare, B Bagchi, N Kumari, B P Mandal, Journal of Mathematical Physics 57, 062106 (2016)
  \bibitem{cs9} A. Mostafazadeh, Phys. Rev. Lett. 102, 220402 (2009).
  \bibitem{o1} Z. H. Musslimani, K. G. Makris, R. El-Ganainy, and D. N. Christodoulides, Phys. Rev. Lett. 100, 030402 (2008).
  
  \bibitem{o3} R. El-Ganainy, K. G. Makris, D. N. Christodoulides and Z. H. Musslimani, Opt. Lett. 32, 2632 (2007).
  \bibitem{o4} A. Guo et al, Phys. Rev. Lett. 103, 093902 (2009).
	
	
	\bibitem{3} H. Raval and U. A. Yajnik, Phys. Rev. D 91, no. 8, 085028 (2015).
	\\ H. Raval and U. A. Yajnik, Springer Proc. Phys. 174, 55 (2016).
	\bibitem{4} Haresh Raval, Eur. Phys. J. C 76:243 (2016).
	\bibitem{ng1} Jean Alexandre, Carl M. Bender, Peter Millington, JHEP 11 (2015) 111
				\bibitem{1} Jean Alexandre, Peter Millington, Dries Seynaeve, Phys. Rev. D 96, 065027 (2017)

	\bibitem{gt}G.'t Hooft, Nucl. Phys. B190, 455 (1981).
			\bibitem{kugo} T. Kugo and I. Ojima, Prog. Theor. Phys. Suppl. 66, 1 (1979).

			\bibitem{2} Jean Alexandre, Peter Millington, Dries Seynaeve, arXiv:1710.01076v1 [hep-th]
			
		\bibitem{Becchi:1974md} C. Becchi, A. Rouet and R. Stora, Commun. Math. Phys. 42 (1975) 127.
		\bibitem{Becchi:1975nq} C. Becchi, A. Rouet and R. Stora, Annals Phys. 98 (1976) 287.
		\bibitem{37} Haresh Raval, B. P. Mandal, Eur. Phys. J. C (2018) 78:416.
		\bibitem{new1} Kei-Ichi Kondo and Toru Shinohara, Phys. Lett. B 491, 263(2000).
		\bibitem{new2}  D. Dudal and H. Verschelde, J. Phys. A36:8507-8516, 2003
		\bibitem{new3} D. Dudal, J. A. Gracey, V. E. R. Lemes, M. S. Sarandy, R. F. Sobreiro, S. P. Sorella, and H. Verschelde, Phys. Rev.	D 70, 114038 (2004)
\bibitem{cd} C. D. Roberts, A. G. Williams, and G. Krein, Int. J. Mod. Phys. A 07, 5607 (1992).
        \bibitem{new4} A. C. Aguilar, D. Binosi, J. Papavassiliou, 	arXiv:1511.08361 [hep-ph].
        \bibitem{new5}  Arlene C. Aguilar and Joannis Papavassiliou, JHEP 12, 012 (2006)
		\bibitem{qed} O.M.Del Cima, Phys. Lett. B  750 (2015) 1–5.
			
			
		\end{thebibliography}
\end{document}